\documentclass[12pt]{article}
\usepackage{graphicx,float,subfigure,epsfig,color,rotating,latexsym}
\usepackage[latin1]{inputenc}
\usepackage[T1]{fontenc}
\def\be{\begin{equation}}
\def\ee{\end{equation}} 
\def\bea{\begin{eqnarray}}
\def\eea{\end{eqnarray}} 
\def\ba{\begin{array}} 
\def\ea{\end{array}}

\def\nin{\noindent}

\begin{document}
\begin{center} 

{\bf \large{Dark energy and Josephson junctions }}\\

\vspace*{0.8 cm}

Vincenzo Branchina\footnote{vincenzo.branchina@ct.infn.it}\label{one}
\vspace*{0.4 cm}

Department of Physics, University of
Catania and \\ INFN, Sezione di Catania, 
Via Santa Sofia 64, I-95123, Catania, Italy 

\vspace*{0.6 cm}

Marco Di Liberto\footnote{madiliberto@ssc.unict.it}\label{two}

\vspace*{0.4 cm}

Scuola Superiore di Catania, Via S. Nullo 5/i, Catania, Italy  

\vspace*{0.6 cm}

Ivano Lodato\footnote{ivlodato@ssc.unict.it}\label{three}

\vspace*{0.4 cm}
Scuola Superiore di Catania, Via S. Nullo 5/i, Catania, Italy and\\
INFN, Sezione di Catania, Via Santa Sofia 64, I-95123, Catania, Italy 

\vspace*{1.2 cm}

{\LARGE Abstract}\\

\end{center}

It has been recently claimed that dark energy can be (and has 
been) observed in laboratory experiments by measuring the power 
spectrum $S_I(\omega)$ of the noise current in a resistively 
shunted Josephson junction and that in new dedicated experiments, 
which will soon test a higher frequency range, $S_I(\omega)$ 
should show a deviation from the linear rising observed in the 
lower frequency region because higher frequencies should not 
contribute to dark energy. Based on previous work on 
theoretical aspects of the fluctuation-dissipation theorem, we 
carefully investigate these issues and show that these claims 
are based on a misunderstanding of the physical origin 
of the spectral function $S_I(\omega)$. According to our 
analysis, dark energy has never been (and will never be) 
observed in Josephson junctions experiments. We also   
predict that no deviation from the linear rising behavior of 
$S_I(\omega)$ will be observed in forthcoming experiments.
Our findings provide new (we believe definite) arguments 
which strongly support previous criticisms.
\vspace*{0.5cm}

\section{Introduction}

The origin of dark energy is one of the greatest mysteries  
confronting theoretical and experimental physics. Different 
proposals for the solution of this so called cosmological 
constant problem are put forward and many review articles nowadays 
discuss and compare these alternative approaches (see for instance 
\,\cite{peeb},\,\cite{padam},\,\cite{cope},\,\cite{nobe}). 

Among many other issues, Copeland et al.\,\cite{cope} 
discuss the suggestion of Beck and Mackey\,\cite{bema1, bema2} according 
to which dark energy can be (and has been) observed in laboratory experiments 
\,\cite{koch} by measuring the power spectrum of the noise current in a 
resistively shunted Josephson junction. If true, 
this would mark a dramatic progress in our understanding of the origin 
of dark energy.

According to\,\cite{bema1, bema2}, this power spectrum is due to thermal
and vacuum fluctuations of the electromagnetic field in the resistor and these 
experiments\,\cite{koch} provide a measurement of the electromagnetic 
zero point energies, which they consider as being at the origin of dark 
energy. 

These ideas have generated a certain debate and some 
authors\,\cite{jetz1, doran, maha} have argued 
against them. Beck and Mackey have rebutted these 
criticisms\,\cite{bema3} but they were again criticized in 
\,\cite{jetz2}. Copeland et al.\,\,close the section of their review devoted 
to this issue by saying that ``time will tell who (if either) are 
correct''\,\cite{cope}. Needless to say, this issue is of the greatest
importance and deserves further investigation.  

Scope of this work is to bring additional elements to the analysis of 
this problem in the hope that the question posed in\,\cite{cope} could 
finally find an answer. To this end, it is necessary to review in some 
detail the Beck and Mackey proposal\,\cite{bema1, bema2}.

These authors begin by considering the work 
of Koch et al.\,\cite{koch}, where the spectral density $S_I(\omega)$ 
of the noise current in the resistor of a resistively shunted Josephson 
junction was measured and confronted against the theoretical prediction, 
\be
\label{spectr}
S_I(\omega) = \frac{4}{R}\left(\frac{\hbar\omega}{2}+
\frac{\hbar\omega}{\exp(\hbar\omega/k_BT)-1)}\right)\,,
\ee
and good agreement was found between experimental results and theory  
($T$ is the temperature and $R$ the resistance of the resistive shunt). 

Eq.\,(\ref{spectr}) comes from an application of the fluctuation-dissipation 
theorem (FDT)\,\cite{cawe} and we immediately recognize the term in parenthesis
as the mean energy of a quantum harmonic oscillator of frequency 
$\omega$ in a thermal bath. Where does this Bose-Einstein (BE) distribution 
factor come from? Does it reflect an underlying 
``harmonic oscillator structure'' of the system\,\cite{taylor}? If yes, 
which harmonic 
oscillators are involved in Eq.\,(\ref{spectr})? A correct answer 
to these questions will turn out to be crucial in understanding the status 
of the Beck and Mackey proposal\,\cite{bema1, bema2} and, we believe, in 
settling the controversy. 

Beck and Mackey interpret this factor as coming from the modes of 
the electromagnetic field in interaction with the charged particles 
\,\cite{bema2} and claim that this experiment provides a 
direct measurement of vacuum fluctuations of the 
electromagnetic field (the $\frac{\hbar\omega}{2}$ term).
Moreover, they assume that dark energy originates from vacuum 
fluctuations of fundamental quantum fields and conjecture that only 
those fluctuations which can be measured in terms of a physical power 
spectrum are gravitationally active, i.e. contribute 
to dark energy. Then, by observing that for strong and electroweak interactions 
it is unlikely that a suitable macroscopic detector exists that can measure the 
corresponding vacuum spectra, they conclude that the only candidate where we know 
that a suitable macroscopic detector exists is the electromagnetic 
interaction\,\cite{bema2}. 

Accordingly, by noting that astrophysical measurements 
give $\varrho_{dark} \sim 10^{-47} GeV^4$ (in natural units), they argue that
there should be a physical cut-off frequency 
$\nu_c =\frac{\omega_c}{2\pi}\sim 1.7\, {\rm THz}$ such that, for frequencies 
above this cut-off, the spectral 
function of the noise current in the Josephson junction should behave
differently than in Eq.\,(\ref{spectr}).  
According to their hypothesis, in fact, for $\omega \geq \omega_c $ the 
$\frac{\hbar\omega}{2}$ term should be absent. 

Coming back to the BE distribution factor in Eq.\,(\ref{spectr}),
Jetzer and Straumann\,\cite{jetz1} (see also \cite{kubo}) 
observed that this term simply comes 
from ratios of Boltzmann factors which appear in the derivation 
of the FDT and stressed that the 
$\frac{\hbar\omega}{2}$ term has nothing to do with zero point energies, 
while Beck and Mackey reply that this is contrary to the view 
commonly expressed in the literature\,\cite{bema3}. 

A simple look to the derivation of the FDT (see Section 2) shows that, as 
for the ratios of Boltzmann factors, Jetzer and Straumann\,\cite{jetz1}
are definitely right. Nevertheless, in a sense that we are going to 
make clear in the following, there is an element of truth in the common 
lore according to which this factor can be regarded as due to a sort of 
underlying harmonic oscillator structure of the system (the resistive 
shunt in the case of the Koch et al. experiment\,\cite{koch}).

In a recent paper\,\cite{noi} we have shown that whenever 
linear response theory applies, which is the main hypothesis under which 
the FDT is derived, any generic bosonic and/or fermionic system can be 
mapped onto a fictitious system of harmonic oscillators in such a manner 
that the quantities appearing in the FDT coincide with the corresponding 
quantities of the fictitious one (for completeness, in sections 2 and 3 we 
briefly review these results. For a comprehensive exposition, however,  
see\,\cite{noi}).  

This allows us to understand {\it in which sense} the harmonic 
oscillator interpretation can be put forward so that we shall be able 
to say whether the Beck and Mackey's proposal is tenable or not. 
We shall see that it is not. 

The rest of the paper is organized as follows. In Section 2 we 
briefly review the derivation of the FDT and consider two convenient 
expressions for the power spectrum of the fluctuating 
observable and for the imaginary part of the corresponding generalized 
susceptibility respectively. In Section 3 we consider the special case 
of a system of harmonic oscillators in interaction with an external field 
and show how the above mentioned mapping is constructed. In Section 4 
we apply the results of the two previous sections to our problem, 
namely the dark energy interpretation\,\cite{bema1,bema2} of the measured 
power spectrum $S_I(\omega)$\,\cite{koch} and show that this interpretation 
is untenable. Section 5 is for our conclusions. 

\section{The fluctuation-dissipation theorem}

In the present section we briefly review the derivation of the 
FDT (see\,\cite{kubo} for more details) and provide expressions for 
the spectral function and the imaginary part of the generalized 
susceptibility which will be useful for our following considerations. 

Consider a macroscopic system with unperturbed 
hamiltonian $\hat{H}_0$ under the influence of the perturbation 
\be
\label{inter}
\hat{V} = - f(t)\,\hat {A}(t)\,,
\ee
where $\hat{A}(t)$ is an observable (a bosonic operator) of 
the system and $f(t)$ an external 
generalized force\footnote{More generally, we could consider 
a local observable and a local generalized force, in which case 
we would have $\hat{V} = -\int d^3\,\vec r \hat{A}(\vec{r})f(\vec{r},t)$, 
and successively define a local susceptibility 
$\chi(\vec{r},t;\vec{r'},t')$ (see Eq.\,(\ref{chi}) below).  
As this would add nothing to our argument, we shall restrict ourselves
to $\vec r$-independent quantities. The extension 
to include local operators is immediate.}. 
Let $|E_n\rangle$ be the $\hat{H}_0$ 
eigenstates (with eigenvalues $E_n$) and  
$\langle E_n|\hat{A}(t)|E_n \rangle =0$. 
Within the framework of linear response theory, the quantum-statistical 
average $\langle\hat{A}(t)\rangle_f$ of the observable $\hat{A}(t)$ 
in the presence of $\hat{V}$ is given by 
\be
\label{resp2}
\langle \hat{A}(t)\rangle_f = \int_{-\infty}^t\, d t' \chi_{_{A}}(t-t') f(t') 
\ee
where $\chi_{_{A}}(t - t')$ is the generalized susceptibility,
\be
\label{chi}
\chi_{_{A}}(t - t')=\frac{i}{\hbar}\theta(t-t') 
\langle [\hat{A}(t),\hat{A}(t')] \rangle = 
-\frac{1}{\hbar}G_R(t - t')\, ,
\ee
with $\langle ... \rangle = 
\sum_{n} \varrho_n \langle E_n| ... |E_n \rangle$,\, 
$\varrho_n= e^{-\beta E_n}/Z$\, ,  $Z=\sum_n e^{- \beta E_n}$\,, 
$G_R(t-t')$ being the retarded Green's function and 
$\hat{A}(t)=e^{i\hat{H_0}t/\hbar}\hat{A}e^{-i\hat{H_0}t/\hbar}$.

If we now consider the mean square of the observable $\hat A(t)$
and write the generalized susceptibility $\chi_{_{A}}$ as 
$\chi_{_{A}} = \chi^{'}_{_{A}} + i\,\chi^{''}_{_{A}}$ (with 
$\chi^{'}_{_{A}}$ the real part and $\chi^{''}_{_{A}}$ the 
imaginary part of $\chi_{_{A}}$), it is not difficult to 
show (see \cite{kubo} and \cite{noi}) that the Fourier 
transform $\langle \hat{A}^2(\omega)\rangle$ of 
$\langle \hat{A}^2(t)\rangle$ is related to the Fourier 
transform $\chi_{_{A}}^{\,''}(\omega)$ through the relation
\be \label{fddt} 
\langle \hat{A}^2(\omega)\rangle = \hbar \chi_{_{A}}^{\,''}(\omega) \, \frac
{1 + e^{- \beta\hbar\omega}}
{1 - e^{- \beta\hbar\omega}} =
\hbar \chi_{_{A}}^{\,''}(\omega) \,{\rm coth}\left( \frac{\beta\hbar\omega}{2}\right)
=2\,\hbar \chi_{_{A}}^{\,''}(\omega) \,
\left(\frac1 2 + \frac{1}{e^{\beta\hbar\omega}-1}\right)\,, 
\ee
which is the celebrated FDT. 

In Eq.\,(\ref{fddt}) we recognize the ratio of Boltzmann factors
alluded by Jetzer and Straumann\,\cite{jetz1}. Actually, it was 
already observed by Kubo et al.\,\cite{kubo} that 
the BE factor in Eq.\,(\ref{fddt}) is simply due to a peculiar 
combination of Boltzmann weights and that there is no reference to 
physical harmonic oscillators of the system whatsoever. 
However, as we already said in the Introduction, there is an 
element of truth in the common lore which considers this term 
as due to a sort of harmonic oscillator structure of 
the system (the resistive shunt in the case 
of the Koch et al.\,\,experiment\,\cite{koch}).

In order to show that, we now refer to our recent work\,\cite{noi}, 
where we have derived 
the following useful expressions for $\langle \hat{A}^2(\omega)\rangle$
and $\chi_{_A}^{\,''}(\omega)$ :
\bea
\langle \hat{A}^2(\omega)\rangle &=& \pi\sum_{j > i}(\varrho_i - \varrho_j) |A_{ij}|^2
\,{\rm coth}\left( \frac{\beta\hbar\omega_{ji}}{2}\right)
\left[\delta\left(\omega -\omega_{ji}\right)+
\delta\left(\omega + \omega_{ji}\right)\right]\label{o5}\\
&=& \pi\,{\rm coth}\left(\frac{\beta\hbar\omega}{2}\right)\sum_{j > i}
(\varrho_i - \varrho_j) |A_{ij}|^2 \left[\delta\left(\omega -\omega_{ji}\right)-
\delta\left(\omega + \omega_{ji}\right)\right]\,,\label{o6}\\
\chi''(\omega)&=&\frac{\pi}{\hbar}\sum_{j > i}(\varrho_i - \varrho_j) |A_{i j}|^2
\left[\delta\left(\omega - \omega_{ji}\right)-
\delta\left(\omega + \omega_{ji}\right)\right]\,\label{chii3}. 
\eea
Clearly, from Eqs.\,(\ref{o6}) and (\ref{chii3}) 
the FDT (Eq.\,(\ref{fddt}))
is immediately recovered. However, what matters for our scopes are the
explicit expressions in Eqs.\,(\ref{o5}) and (\ref{chii3}). Starting 
from these equations, in fact, we can easily show that it is possible
to build up a mapping between the real system and a fictitious system
of harmonic oscillators\,\cite{noi} in such a manner that 
$\chi_{_A}^{\,''}(\omega)$ and $\langle \hat{A}^2(\omega)\rangle$
are exactly reproduced by considering the corresponding quantities of the 
fictitious system. In the following section we outline the main steps for 
this construction (see \cite{noi} for details).

\section{The Mapping}
In order to build up this mapping, we consider first a system 
${\cal S}_{osc}$ of 
harmonic oscillators (each of which is labeled below by the double 
index $\{ji\}$ for reasons that will become clear in the following) 
whose free hamiltonian is: 
\be\label{armonico}
\hat H_{osc} = \sum_{j > i}\left(\frac{\hat p_{ji}^{\,2} }{2 M_{ji}} + 
\frac{M_{ji} \omega_{ji}^2}{2}\,\hat q_{ji}^{\,2}\right)\,,
\ee
where $\omega_{ji}$ are the proper 
frequencies of the individual harmonic oscillators and $M_{ji}$ their 
masses. Let  $| n_{j i}\rangle$ ($n_{ji}=0, 1,2,...$) be the 
occupation number states of the $\{ji\}$ oscillator out of which the Fock 
space of ${\cal S}_{osc}$ is built up.
Let us consider also ${\cal S}_{osc}$ in interaction with an external 
system through the one-particle operator: 
\be\label{armint}
\hat V_{osc} = - f(t) \hat A_{osc}\,, 
\ee
with 
\be\label{onepart}
{\hat A}_{osc} = \sum_{j > i} \left(\alpha_{j i} \,{\hat q}_{ji} \right)\,.
\ee
Obviously, the FDT applied to ${\cal S}_{osc}$ gives\, 
$\langle {\hat A}_{osc}^2(\omega)\rangle = 
\hbar \chi_{osc}^{\,''}(\omega) \,{\rm coth}\left( \frac{\beta\hbar\omega}{2}\right)$\,,
but this is not what matters to us. 

What is important for our purposes is that, as shown in\,\cite{noi},
for ${\cal S}_{osc}$ we can exactly compute 
$\langle {\hat A}_{osc}^2(\omega) \rangle$ and $\chi_{osc}^{\,''}(\omega)$. 
The reason is that for this system, differently from any other generic 
system, we can explicitly compute the matrix elements of ${\hat A}_{osc}$. 
The result is (compare with Eqs.\,(\ref{o5}), (\ref{o6}) and (\ref{chii3})): 
\bea
\langle \hat{A}_{osc}^2(\omega)\rangle&=&\pi
\sum_{j>i} \, \alpha_{ji}^2\, \frac{\hbar}{2 M_{ji} \omega_{ji}}\,
{\rm coth}\left(\frac{\beta \hbar\omega_{ji}}{2}\right)
[\delta (\omega-\omega_{ji}) +\delta (\omega+\omega_{ji})]\label{osc5}\\
&=& \pi\, {\rm coth}\left(\frac{\beta \hbar\omega}{2}\right)\sum_{j>i} \,
\alpha_{ji}^2\,\frac{\hbar}{2 M_{ji} \omega_{ji}}\,
[\delta (\omega-\omega_{ji}) - \delta (\omega+\omega_{ji})]\label{oscc5}\,;\\
\chi_{osc}^{''} (\omega)&=&\frac{\pi}{\hbar}
\sum_{j>i} \,\alpha_{ji}^2\,\frac{\hbar}{2 M_{ji} \omega_{ji}}\,
[\delta (\omega-\omega_{ji}) - \delta (\omega+\omega_{ji})]\,\label{chi3}.
\eea

Naturally, comparing Eq.\,(\ref{oscc5}) with Eq.\,(\ref{chi3}) we 
see that for ${\cal S}_{osc}$ the FDT holds true, 
as it should. However, for our scopes it is important to note the
following. For this system, the 
${\rm coth}\left(\frac{\beta \hbar\omega}{2}\right)$ factor of the FDT 
originates from the {\it individual contributions}  
${\rm coth}\left(\frac{\beta \hbar\omega_{ji}}{2}\right)$ of each of 
the harmonic oscillators of ${\cal S}_{osc}$.

We can now build up our mapping. 
Let us consider the original system ${\cal S}$, described by the 
unperturbed hamiltonian $\hat H_0$\,, in interaction with an external 
field $f(t)$ through the interaction term  $\hat V = - f (t)\,\hat A$  
(see Eq.\,(\ref{inter})), and construct a fictitious system of harmonic 
oscillators ${\cal S}_{osc}$, described by the free hamiltonian 
${\hat H}_{osc}$ of Eq.\,(\ref{armonico}), in interaction with the same 
external field $f(t)$ through the interaction term ${\hat V}_{osc}$ of 
Eq.\,(\ref{armint}), with $\hat A_{osc}$ given by Eq.\,(\ref{onepart}),  
where for $\alpha_{j i}$ we choose
\be\label{alfa}
\alpha_{j i} = \left(\frac{2 M_{ji} \omega_{ji}}{\hbar}\right)^{\frac12}
(\varrho_i - \varrho_j)^{\frac12} \,|A_{ij}|
\ee
and for the proper frequencies $\omega_{ji}$ of the oscillators 
\be\label{omega}
\omega_{ji}= (E_j-E_i)/\hbar > 0\,,
\ee
with $E_i$ the eigenvalues of the hamiltonian ${\hat H}_0$ of the real system. 

By comparing Eq.\,(\ref{oscc5}) with Eq.\,(\ref{o6}) and Eq.\,(\ref{chi3}) 
with Eq.\,(\ref{chii3}), it is immediate to see that 
with the above choices of $\alpha_{ji}$ and $\omega_{ji}$ we have: 
\bea
\langle \hat{A}^2(\omega)\rangle &=& 
\langle \hat{A}_{osc}^2(\omega)\rangle\label{cen1}\\
\chi_{_A}^{''} (\omega) &=& \chi_{osc}^{''} (\omega)\label{cen2}\,.
\eea
Eqs.\,(\ref{cen1}) and (\ref{cen2}) define the mapping we are looking for. 
They show that it is possible to map the real 
system ${\cal S}$ onto a fictitious system of harmonic oscillators
${\cal S}_{osc}$\,, 
\be
{\cal S} \,\, \to \,\, {\cal S}_{osc}\,,
\ee
in such a manner that $\chi_{_A}^{''} (\omega)$ and 
$\langle \hat{A}^2(\omega)\rangle$ of the real system
are equivalently obtained by computing the 
corresponding quantities of the fictitious one.
The key ingredient to construct such a 
mapping is the hypothesis that linear response theory is applicable 
(which is the central hypothesis under which the FDT is established). 

Now, by considering the ``equivalent'' harmonic oscillators system  
${\cal S}_{osc}$ rather than the real one, we can somehow  
regard the BE distribution factor 
${\rm coth}\left( \frac{\beta\hbar\omega}{2}\right)$ 
of the FDT in Eq.\,(\ref{fddt}) as originating from the individual
contributions ${\rm coth}\left( \frac{\beta\hbar\omega_{ji}}{2}\right)$ 
of each of the oscillators of the equivalent fictitious system (see above,
Eqs.\,(\ref{osc5}), (\ref{oscc5}) and (\ref{chi3})). 
In this sense, this mapping allows for an oscillator 
interpretation of the BE term in the FDT.

At the same time, however, the above findings clearly teach us
that the BE distribution term in the FDT {\it does not describe the 
physical nature of the system}. It rather encodes a fundamental property of 
any bosonic and/or fermionic system: whenever 
linear response theory is applicable, any generic system is equivalent 
(in the sense defined above) to a system of quantum harmonic oscillators.

\section{Dark energy and laboratory experiments}

We are now in the position to apply the results of the two previous 
sections to our problem. As we said in the Introduction, Beck 
and Mackey\,\cite{bema1, bema2} interpret the Koch et al.\,experimental 
results\,\cite{koch} for the spectral density $S_I(\omega)$ of the 
noise current in a resistively shunted Josephson junction 
as a direct measurement of {\it vacuum fluctuations 
of the electromagnetic field} in the shunt resistor. Moreover, according 
to their ideas, these zero-point energies are nothing but the dark energy 
of the universe. 

In view of our results, however, this interpretation seems to be 
untenable. Eq.\,(\ref{spectr}) for $S_I(\omega)$ comes from an 
application of the FDT to the case of the noise current in the 
shunt resistor. Therefore, according to our findings,
the BE distribution factor which appears in $S_I(\omega)$ 
has nothing to do with thermal and vacuum fluctuations of the 
electromagnetic field in the resistor. Our analysis shows that this 
factor rather reflects a general property of any quantum system 
valid whenever linear response theory applies. The resistor 
(as well as any other generic system) can be mapped onto a system 
of fictitious harmonic oscillators in such a manner that the 
the power spectrum of the noise current and the related susceptibility 
can be reproduced by considering the equivalent quantities for 
the fictitious oscillators. 
 
It is in this sense, and {\it only in this sense}, that
the BE factor can be interpreted in terms of harmonic oscillators, 
no other physical meaning can be superimposed on it. 
According to these considerations, we conclude that 
the claim that dark energy is 
observed in laboratory experiments\,\cite{bema1,bema2} is 
based on an incorrect interpretation of the origin of the BE 
factor in the FDT.

We believe that this should help in solving  
the controversy, which is left open in the Copeland 
et al.\,review\,\cite{cope}, between the 
proponents\,\cite{bema1, bema2} of the dark energy interpretation 
of the Koch et al.\,experiments\,\cite{koch} and the 
opponents\,\cite{jetz1, doran, maha, jetz2}.  
In this respect, it is worth to stress that our analysis provides 
new arguments which strongly support the conclusions 
of these latter works\,\cite{jetz1, doran, maha, jetz2}.  

A distinctive new element of our work, which in our opinion 
should greatly help in settling the question, concerns the 
interpretation of the FDT presented in section 5 
of\,\cite{bema2}. These authors note that, although the FDT is valid 
for arbitrary hamiltonians $H$, where $H$ need not to describe 
harmonic oscillators, in the FDT appears a {\it universal function}
$H_{uni}$,  
$H_{uni}= \frac12\hbar\omega +\hbar\omega/(exp(\hbar\omega/kT)-1)$,
which can always be interpreted as the mean energy of a 
harmonic oscillator. Then, they identify the $\frac12\hbar\omega$ in 
$H_{uni}$ as the source of dark energy. 

The distinctive feature of our analysis is that, with the help of the 
formal mapping discussed in the previous section, which is valid for 
{\it any generic system}, the reason for the appearance of this 
{\it universal} function is immediately apparent. At the same time, 
however, this clearly shows that it cannot be claimed that the Koch 
et al.\,\cite{koch} experimental device is measuring zero point energies. 
As already noted in\,\cite{jetz1,jetz2}, these experiments simply 
measure a general quantum property of the system, the $\frac12\hbar\omega$ 
in $S_I(\omega)$ has nothing to do with zero point energies.

Another very important point related to these issues concerns 
future measurements\,\cite{barb,warb} of the power spectrum 
$S_I(\omega)$ for values 
of the frequency higher than those measured by Koch et al.\,\cite{koch}.  
In fact, according to Beck and Mackey\,\cite{bema1,bema2}, in 
forthcoming experiments\,\cite{barb,warb}, which are purposely 
designed to test a higher frequency range of $S_I(\omega)$, we should 
observe a dramatic change in the 
behavior of the spectral function $S_I(\omega)$ for these higher 
values of the frequency due to the presence 
of a cut-off which separates the gravitationally active modes from 
those which are not gravitationally active (see the Introduction). 
In view of our findings, however, we do not expect to observe in these 
experiments\,\cite{barb,warb} any change in the behavior of 
$S_I(\omega)$. We simply state that such a cut-off does not exist.

In this respect, we note that Beck and Mackey have recently proposed a 
new model for dark energy which should naturally incorporate such a 
cut-off\,\cite{bm3}. According to our analysis, this model 
seems to be deprived of any experimental and theoretical support.

As a consequence of our results, a deviation of 
$S_I(\omega)$ from the behavior given in Eq.\,(\ref{spectr}) 
could be observed only if the central hypothesis on which the 
derivation of the FDT is based, namely the applicability of linear 
response theory, no longer holds true in this higher frequency region.

\section{Summary and Conclusions}
With the help of a general theorem, which shows that (under the 
assumption that linear response theory is applicable) any bosonic 
and/or fermionic fermionic  
system can be mapped onto a fictitious system of harmonic oscillators,
we have shown that the appearance of a Bose-Einstein distribution 
factor in the power spectrum of the noise current of a resistively 
shunted Josephson junction\,\cite{koch} has nothing to do with a 
real (physical) harmonic oscillator structure of the shunt resistor. 
We then conclude that, contrary to recent claims\,\cite{bema1, bema2}, 
experiments where this power spectrum was measured\,\cite{koch} do not 
provide any direct measurement of zero point energies and, as a 
consequence, no dark energy has ever been measured in these laboratory 
experiments. 

A direct consequence of our analysis is that, contrary to what is 
predicted in\,\cite{bema1, bema2}, we do not expect any 
deviation from the linear rising behavior of $S_I(\omega)$ with $\omega$. 
According to our analysis, in fact, the $\frac12\hbar\omega$ term in 
$S_I(\omega)$ has nothing to do with the dark energy in the universe,  
therefore we do not expect any cut-off which separates the gravitationally 
active zero point energies from the gravitationally non-active ones.

Finally, our analysis suggests that the theory which should naturally 
incorporate such a cut-off\,\cite{bm3} is deprived of any experimental 
and theoretical foundation.

We believe that our work provides a satisfactory answer to the intriguing 
and important question left open by Copeland et al. in the section of their 
review devoted to the possibility of measuring dark energy in laboratory 
experiments\,\cite{cope}: ``time will tell who (if either) are correct''.
According to our analysis, the opponents to the dark energy interpretation 
of the Koch et al.\,\,experiments\,\cite{koch} are correct.

\vfill\eject
\nin
{\large{\bf Acknowledgments}}
\vskip 6pt
We would like to thank Luigi Amico, Marcello Baldo, Pino Falci and 
Dario Zappal\`a for many useful discussions.

\end{document}